%% file: main.tex
\documentclass{article}

\usepackage[margin=1in]{geometry}

\usepackage[comma]{natbib}

\usepackage[utf8]{inputenc}
\usepackage[T1]{fontenc}
\usepackage[english]{babel}
\usepackage{graphicx}
\usepackage{xcolor}
\usepackage{hyperref}
\graphicspath{{images/}{../images/}}
\usepackage{subfiles}
\usepackage{subcaption}
\usepackage{wrapfig}
\usepackage[utf8]{inputenc}
\usepackage{graphicx}
\usepackage{comment}
\usepackage{amssymb}
\usepackage{amsmath}

\DeclareMathOperator*{\argmin}{arg\,min}
\usepackage{changepage}
\usepackage[noblocks]{authblk}

\newcommand\cmnt[2]{{\textcolor{red}{[{\em #1 --- #2}] }}}
\newcommand\franks[1]{\cmnt{#1}{Franks}}
\newcommand\cmntt[2]{{\textcolor{blue}{[{\em #1 --- #2}] }}}
\newcommand\Terner[1]{\cmntt{#1}{Terner}}
 
\begin{document}

\markboth{Zachary Terner and Alexander Franks}{Basketball Analytics}

% Title
\title{Modeling Player and Team Performance in Basketball}

%Authors, affiliations address.
\author[1]{Zachary Terner}
\author[2]{Alexander Franks}

\affil[1]{Department of Statistics and Applied Probability, University of California Santa Barbara, Santa Barbara, CA, 93106;  zterner@ucsb.edu}
\affil[2]{Department of Statistics and Applied Probability, University of California Santa Barbara, Santa Barbara, CA, 93106; afranks@pstat.ucsb.edu}

\maketitle
%Abstract
\begin{abstract}
%A rise in data availability -- including tracking data, where players and the ball are tracked through space and time -- has led to the development of new statistical methods for understanding the game of basketball. 

In recent years, analytics has started to revolutionize the game of basketball: quantitative analyses of the game inform team strategy, management of player health and fitness, and how teams draft, sign, and trade players.  In this review, we focus on methods for quantifying and characterizing basketball gameplay. At the team level, we discuss methods for characterizing team strategy and  performance, while at the player level, we take a deep look into a myriad of tools for player evaluation. This includes metrics for overall player value, defensive ability, and shot modeling, and methods for understanding performance over multiple seasons via player production curves. We conclude with a discussion on the future of basketball analytics, and in particular highlight the need for causal inference in sports.

\end{abstract}

 %\begin{keywords}
%Basketball analytics, hierarchical modeling, spatio-temporal data, Markov chains, regularization
 %\end{keywords}

%Table of Contents
%\tableofcontents 

 \pagebreak
% \section{Outline}
 
% \subfile{outline}
 
%  \textit{The Editorial Committee encourages you to provide an overview of your chosen topic in the article’s Introduction to entice the uninitiated and to present the major unsolved issues, which you may then discuss more fully in the concluding section and list of Future Issues. We hope that you will present a critical discussion of the current status of the field, rather than an encyclopedic coverage of many papers. Given the breadth of statistical advances and the inability of textbooks to remain current, we suggest reviewing your topic with a broad audience in mind, including both experts and those new to the field. We also welcome your personal perspective, especially with respect to what you think is most important and where the field is going, yet the presentation must be balanced. In short, we want a compelling and up-to-date presentation that communicates the opportunities and excitement of the subject.}

\section{Introduction}

% These are all spatial
% need to cite \citep{ripley2005spatial, diggle2013statistical} (removed entries that were cited)
% cressie93 (statistics for spatial?)
% 

Basketball is a global and growing sport with interest from fans of all ages. This growth has coincided with a rise in data availability and innovative methodology that has inspired fans to study basketball through a statistical lens. Many of the approaches in basketball analytics can be traced to pioneering work in baseball~\citep{schwartz_2013}, beginning with Bill James' publications of \emph{The Bill James Baseball Abstract} and the development of the field of ``sabermetrics''~\citep{james1984the-bill, james1987bill, james2010new}. James' sabermetric approach captivated the larger sports community when the 2002 Oakland Athletics used analytics to win a league-leading 102 regular season games despite a prohibitively small budget. Chronicled in Michael Lewis' \textit{Moneyball}, this story demonstrated the transformative value of analytics in sports~\citep{lewis2004moneyball}. 
 
In basketball, Dean Oliver and John Hollinger were early innovators who argued for evaluating players on a per-minute basis rather than a per-game basis and developed measures of overall player value, like Hollinger's Player Efficiency Rating (PER)~\citep{oliver2004basketball, hollingerper, hollinger2004pro}. The field of basketball analytics has expanded tremendously in recent years, even extending into popular culture through books and articles by data-journalists like Nate Silver and Kirk Goldsberry, to name a few~\citep{silver2012signal, goldsberry2019sprawlball}. In academia, interest in basketball analytics transcends the game itself, due to its relevance in fields such as psychology \citep{gilovich1985hot, vaci2019large, price2010racial}, finance and gambling \citep{brown1993fundamentals, gandar1998informed}, economics (see, for example, the Journal of Sports Economics), and sports medicine and health \citep{drakos2010injury, difiori2018nba}.  
 
Sports analytics also has immense value for statistical and mathematical pedagogy.  For example, \citet{drazan2017sports} discuss how basketball can broaden the appeal of math and statistics across youth.  At more advanced levels, there is also a long history of motivating statistical methods using examples from sports, dating back to techniques like shrinkage estimation \citep[e.g.][]{efron1975data} up to the emergence of modern sub-fields like deep imitation learning for multivariate spatio-temporal trajectories \citep{le2017data}. Adjusted plus-minus techniques (Section \ref{reg-section}) can be used to motivate important ideas like regression adjustment, multicollinearity, and regularization \citep{sill2010improved}.

\subsection{This review}
 
Our review builds on the early work of \citet{kubatko2007starting} in ``A Starting Point for Basketball Analytics,'' which aptly establishes the foundation for basketball analytics. In this review, we focus on modern statistical and machine learning methods for basketball analytics and highlight the many developments in the field since their publication nearly 15 years ago. Although we reference a broad array of techniques, methods, and advancements in basketball analytics, we focus primarily on understanding team and player performance in gameplay situations. We exclude important topics related to drafting players~\citep[e.g.][]{, mccann2003illegal,groothuis2007early,berri2011college,arel2012NBA}, roster construction, win probability models, tournament prediction~\citep[e.g.][]{brown2012insights,gray2012comparing,lopez2015building, yuan2015mixture, ruiz2015generative, dutta2017identifying, neudorfer2018predicting}, and issues involving player health and fitness~\citep[e.g.][]{drakos2010injury,mccarthy2013injury}.  We also note that much of the literature pertains to data from the National Basketball Association (NBA). Nevertheless, most of the methods that we discuss are relevant across all basketball leagues; where appropriate, we make note of analyses using non-NBA data.

We assume some basic knowledge of the game of basketball, but for newcomers, \url{NBA.com} provides a useful glossary of common NBA terms~\citep{nba_glossary}.  We begin in Section~\ref{datatools} by summarizing the most prevalent types of data available in basketball analytics. The online supplementary material highlights various data sources and software packages. In Section~\ref{teamsection} we discuss methods for modeling team performance and strategy. Section~\ref{playersection} follows with a description of models and methods for understanding player ability. We conclude the paper with a brief discussion on our view on the future of basketball analytics. %% not sure why this is commented out? 

\subsection{Data and tools}
\label{datatools}

\noindent \textbf{Box score data:} The most available datatype is box score data. Box scores, which were introduced by Henry Chadwick in the 1900s~\citep{pesca_2009}, summarize games across many sports. In basketball, the box score includes summaries of discrete in-game events that are largely discernible by eye: shots attempted and made, points, turnovers, personal fouls, assists, rebounds, blocked shots, steals, and time spent on the court. Box scores are referenced often in post-game recaps.  

\url{Basketball-reference.com}, the professional basketball subsidiary of \url{sports-reference.com}, contains preliminary box score information on the NBA and its precursors, the ABA, BAA, and NBL, dating back to the 1946-1947 season; rebounds first appear for every player in the 1959-60 NBA season \citep{nbaref}. There are also options for variants on traditional box score data, including statistics on a per 100-possession, per game, or per 36-minute basis, as well as an option for advanced box score statistics. Basketball-reference additionally provides data on the WNBA and numerous international leagues. Data on further aspects of the NBA are also available, including information on the NBA G League, NBA executives, referees, salaries, contracts, and payrolls as well as numerous international leagues.  One can  find similar college basketball information on the \url{sports-reference.com/cbb/} site, the college basketball subsidiary of \url{sports-reference.com}. 

For NBA data in particular, \url{NBA.com} contains a breadth of data beginning with the 1996-97 season~\citep{nbastats}. This includes a wide range of summary statistics, including those based on tracking information, a defensive dashboard, ''hustle''-based statistics, and other options. \url{NBA.com} also provides a variety of tools for comparing various lineups, examining on-off court statistics, and measuring individual and team defense segmented by shot type, location, etc. The tools provided include the ability to plot shot charts for any player on demand.

\hfill

\noindent \textbf{Tracking data}: Around 2010, the emergence of ``tracking data,'' which consists of spatial and temporally referenced player and game data, began to transform basketball analytics. Tracking data in basketball fall into three categories: player tracking, ball tracking, and data from wearable devices. Most of the basketball literature that pertains to tracking data has made use of optical tracking data from SportVU through Stats, LLC and Second Spectrum, the current data provider for the NBA. Optical data are derived from raw video footage from multiple cameras in basketball arenas, and typically include timestamped $(x, y)$ locations for all 10 players on the court as well as $(x, y, z)$ locations for the basketball at over 20 frames per second.\footnote{A sample of SportVU tracking data can currently be found on Github \citep{github-tracking}.} Many notable papers from the last decade use tracking data to solve a range of problems: evaluating defense \citep{franks2015characterizing}, constructing a ``dictionary'' of play types \citep{miller2017possession}, evaluating expected value of a possession \citep{cervonepointwise}, and constructing deep generative models of spatio-temporal trajectory data \citep{yu2010hidden, yue2014learning, le2017data}.  See \citet{bornn2017studying} for a more in-depth introduction to methods for player tracking data.  

Recently, high resolution technology has enabled $(x,y,z)$ tracking of the basketball to within one centimeter of accuracy. Researchers have used data from NOAH~\citep{noah} and RSPCT~\citep{rspct}, the two largest providers of basketball tracking data, to study several aspects of shooting performance~\citep{marty2018high, marty2017data, bornn2019using, shah2016applying, harmon2016predicting}, see Section \ref{sec:shot_efficiency}. Finally, we also note that many basketball teams and organizations are beginning to collect biometric data on their players via wearable technology. These data are generally unavailable to the public, but can help improve understanding of player fitness and motion~\citep{smith_2018}. Because there are few publications on wearable data in basketball to date, we do not discuss them further.

\hfill

\noindent \textbf{Data sources and tools:} For researchers interested in basketball, we have included two tables in the supplementary material. Table 1 contains a list of R and Python packages developed for scraping basketball data, and Table 2 enumerates a list of relevant basketball data repositories.

%  \franks{Deep learning papers-- no real insights in these papers in my opinion-- where/how can we reference these? Maybe say something about tracking data and deep learning. }

% \citep{harmon2016predicting}“given the trajectories of the players and ball in the previous five seconds, can we accurately predict the likelihood that a player with role X will make the shot?” Where does this paper go? Maybe in intro? .  

\section{Team performance and strategy}
\label{teamsection}

% We begin with a brief discussion on predictors of team success, and finish with an in-depth discussion on models for characterizing team play, strategy and the importance of spacing.  

% \subsection{Predictors of team success}

Sportswriters often discuss changes in team rebounding rate or assist rate after personnel or strategy changes, but these discussions are rarely accompanied by quantitative analyses of how these changes actually affect the team's likelihood of winning. Several researchers have attempted to address these questions by investigating which box score statistics are most predictive of team success, typically with regression models \citep{hofler2006efficiency, melnick2001relationship, malarranha2013dynamic, sampaio2010effects}.  Unfortunately, the practical implications of such regression-based analyses remains unclear, due to two related difficulties in interpreting predictors for team success: 1) multicollinearity leads to high variance estimators of regression coefficients~\citep{ziv2010predicting} and 2) confounding and selection bias make it difficult to draw any causal conclusions. In particular, predictors that are correlated with success may not be causal when there are unobserved contextual factors or strategic effects that explain the association (see Figure \ref{fig:simpsons} for an interesting example).  More recent approaches leverage spatio-temporal data to model team play within individual possessions. These approaches, which we summarize below, can lead to a better understanding of how teams achieve success.

 \label{sec:team}
\subsection{Network models}

One common approach to characterizing team play involves modeling the game as a network and/or modeling transition probabilities between discrete game states. For example, \citet{fewell2012basketball} define players as nodes and ball movement as edges and compute network statistics like degree and flow centrality across positions and teams. They differentiate teams based on the propensity of the offense to either move the ball to their primary shooters or distribute the ball unpredictably.~\citet{fewell2012basketball} suggest conducting these analyses over multiple seasons to determine if a team's ball distribution changes when faced with new defenses.~\citet{xin2017continuous} use a similar framework in which players are nodes and passes are transactions that occur on edges. They use more granular data than \citet{fewell2012basketball} and develop an inhomogeneous continuous-time Markov chain to accurately characterize players' contributions to team play.

% estimates of players' abilities to score, rebound, and steal the ball. 

% However, the authors use more granular data than \citet{fewell2012basketball}, modeling each basketball play as an inhomogeneous continuous-time Markov chain. 
% Their continuous time stochastic block model (CSBM)  can be used to clusters players into novel position types. This analysis hints at the existence of new player positions and can also be used to identify cluster-specific estimates of players' abilities to score, rebound, and steal the ball. 

\citet{skinner2015method} motivate their model of basketball gameplay with a traffic network analogy, where possessions start at Point A, the in-bounds, and work their way to Point B, the basket. With a focus on understanding the efficiency of each pathway, Skinner proposes that taking the highest percentage shot in each possession may not lead to the most efficient possible game. He also proposes a mathematical justification of the ``Ewing Theory'' that states a team inexplicably plays better when their star player is injured or leaves the team~\citep{simmons}, by comparing it to a famous traffic congestion paradox~\citep{skinner2010price}. See \citet{skinner2015optimal} for a more thorough discussion of optimal strategy in basketball.

\subsection{Spatial perspectives}

Many studies of team play also focus on the importance of spacing and spatial context.~\citet{metulini2018modelling} try to identify spatial patterns that improve team performance on both the offensive and defensive ends of the court. The authors use a two-state Hidden Markov Model to model changes in the surface area of the convex hull formed by the five players on the court. The model describes how changes in the surface area are tied to team performance, on-court lineups, and strategy.~\citet{cervone2016NBA} explore a related problem of assessing the value of different court-regions by modeling ball movement over the course of possessions. 
Their court-valuation framework can be used to identify teams that effectively suppress their opponents' ability to control high value regions.
% They propose a definition for a player's ``portfolio value'' as the total value of the space owned by a player and use this to characterize team and player impact. 

Spacing also plays a crucial role in generating high-value shots. ~\citet{lucey2014get} examined almost 20,000 3-point shot attempts from the 2012-2013 NBA season and found that defensive factors, including a ``role swap'' where players change roles, helped generate open 3-point looks. 
% However, they did not find any offensive factors -- including dribbles, passes, etc -- that were predictive of getting an open shot.
% On the surface, this appears to conflict with
In related work, \citet{d2015move} stress the importance of ball movement in creating open shots in the NBA. They show that ball movement adds unpredictability into offenses, which can create better offensive outcomes. The work of D'Amour and Lucey could be reconciled by recognizing that unpredictable offenses are likely to lead to ``role swaps'', but this would require further research.~\citet{sandholtz2019measuring} also consider the spatial aspect of shot selection by quantifying a team's ``spatial allocative efficiency,'' a measure of how well teams determine shot selection. They use a Bayesian hierarchical model to estimate player FG\% at every location in the half court and compare the estimated FG\% with empirical field goal attempt rates. In particular, the authors identify a proposed optimum shot distribution for a given lineup and compare the true point total with the proposed optimum point total. Their metric, termed Lineup Points Lost (LPL), identifies which lineups and players have the most efficient shot allocation.

\begin{figure}
    \centering
    \includegraphics[width=0.85\textwidth]{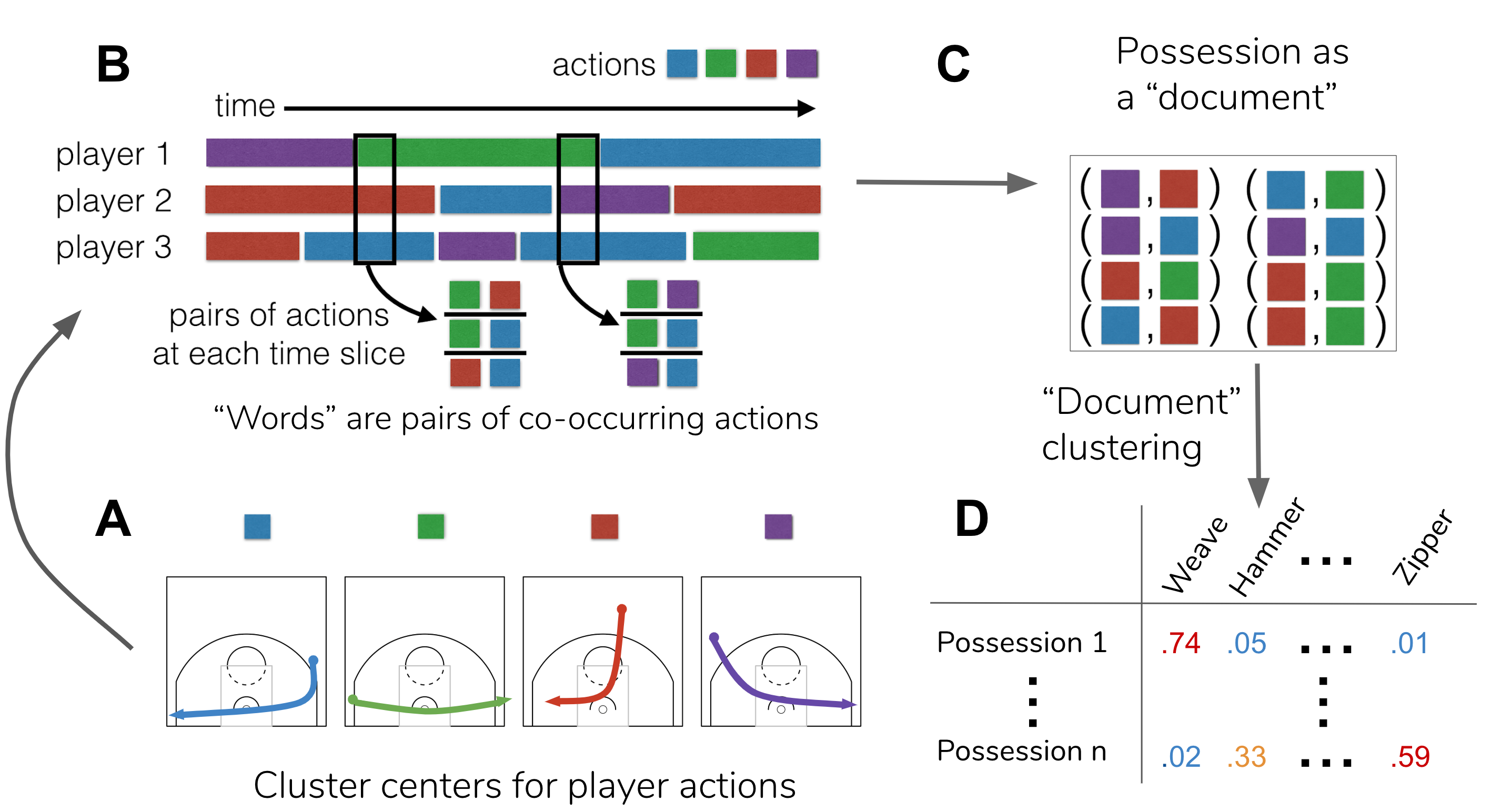}
    \caption{Unsupervised learning for play discovery \citep{miller2017possession}.  A) Individual player actions are clustered into a set of discrete actions. Cluster centers are modeled using Bezier curves.  B) Each possession is reduced to a set of co-occurring actions.  C) By analogy, a possession can be thought of as a ``document'' consisting of ``words.'' ``Words'' correspond to all pairs of co-occurring actions. A ``document'' is the possession, modeled using a bag-of-words model.  D) Possessions are clustered using Latent Dirichlet Allocation (LDA).  After clustering, each possession can be represented as a mixture of strategies or play types (e.g. a ``weave'' or ``hammer'' play).} 
    \label{fig:playbook}
\end{figure}

\subsection{Play evaluation and detection}

Finally, \citet{lamas2015modeling} examine the interplay between offensive actions, or space creation dynamics (SCDs), and defensive actions, or space protection dynamics (SPDs). In their video analysis of six Barcelona F.C. matches from Liga ACB, they find that setting a pick was the most frequent SCD used but it did not result in the highest probability of an open shot, since picks are most often used to initiate an offense, resulting in a new SCD. Instead, the SCD that led to the highest proportion of shots was off-ball player movement. They also found that the employed SPDs affected the success rate of the SCD, demonstrating that offense-defense interactions need to be considered when evaluating outcomes. 

Lamas' analysis is limited by the need to watch games and manually label plays. Miller and Bornn address this common limitation by proposing a method for automatically clustering possessions using player trajectories computed from optical tracking data~\citep{miller2017possession}.  First, they segment individual player trajectories around periods of little movement and use a functional clustering algorithm to cluster individual segments into one of over 200 discrete actions. They use a probabilistic method for clustering player trajectories into actions, where cluster centers are modeled using Bezier curves. These actions serve as inputs to a probabilistic clustering model at the possession level. For the possession-level clustering, they propose Latent Dirichlet Allocation (LDA), a common method in the topic modeling literature~\citep{blei2003latent}. LDA is traditionally used to represent a document as a mixture of topics, but in this application, each possession (``document'') can be represented as a mixture of strategies/plays (``topics''). Individual strategies consist of  a set of co-occurring individual actions (``words''). The approach is summarized in Figure \ref{fig:playbook}.  This approach for unsupervised learning from possession-level tracking data can be used to characterize plays or motifs which are commonly used by teams.  As they note, this approach could be used to ``steal the opponent's playbook'' or automatically annotate and evaluate the efficiency of different team strategies. Deep learning models \citep[e.g.][]{le2017data, shah2016applying} and variational autoencoders could also be effective for clustering plays using spatio-temporal tracking data.

It may also be informative to apply some of these techniques to quantify differences in strategies and styles around the world. For example, although the US and Europe are often described as exhibiting different styles~\citep{hughes_2017}, this has not yet been studied statistically. Similarly, though some lessons learned from NBA studies may apply to The EuroLeague, the aforementioned conclusions about team strategy and the importance of spacing may vary across leagues.

% Additionally, since each possession is summarized as a sequence of strategies, this framework lets one identify similar possessions~\citep{miller2017possession}.  

\begin{comment}
\citep{fichman2018three, fichman2019optimal}
\citep{kozar1994importance} Importance of free throws at different stages of games.  \franks{better here than in the individual performance section?} \Terner{will take a look at the papers and see where they belong}

% \subsubsection{Understanding shot selection strategies}

\citep{ervculj2015basketball} also use hierarchical multinomial logistic regression, to explore differences in shot types across multiple levels of play, across multiple levels of play.

\Terner{This paper looks like it would be a good inclusion in the paper (sandholtz and bornn) but not sure where it fits:}
\citep{sandholtz2018transition} Game-theoretic approach to strategy
\end{comment}

 \section{Player performance}
 \label{playersection}
 
In this section, we focus on methodologies aimed at characterizing and quantifying different aspects of individual performance.  These include metrics which reflect both the overall added value of a player and specific skills like shot selection, shot making, and defensive ability.  
 
\begin{figure}
    \centering
    \includegraphics[width=0.85\textwidth]{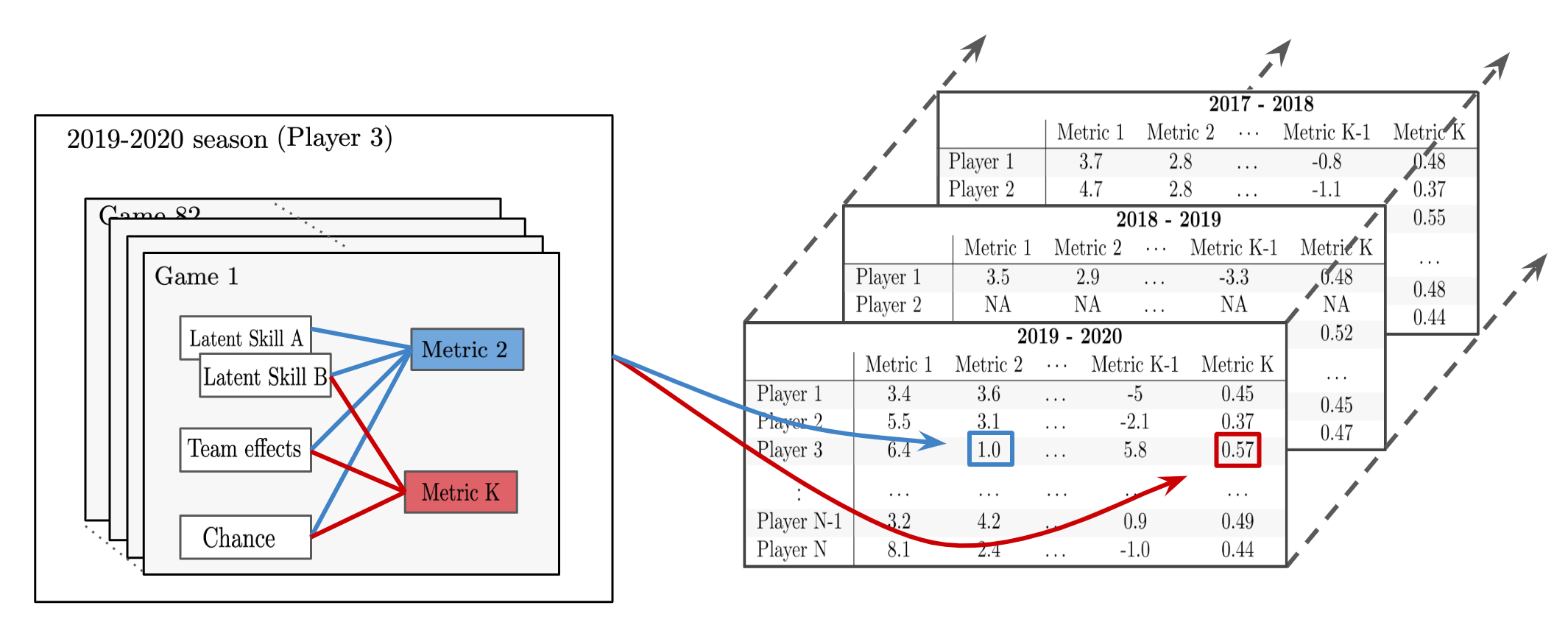}
    \caption{Diagram of the sources of variance in basketball season metrics. Metrics reflect multiple latent player attributes but are also influenced by team ability, strategy, and chance variation.  Depending on the question, we may be interested primarily in differences between players, differences within a player across seasons, and/or the dependence between metrics within a player/season. Player 2 in 2018-2019 has missing values (e.g. due to injury) which emphasizes the technical challenge associated with irregular observations and/or varying sample sizes.}
    \label{fig:player_variance}
\end{figure}

When analyzing player performance, one must recognize that variability in metrics for player ability is driven by a combination of factors.  This includes sampling variability, effects of player development, injury, aging, and changes in strategy (see Figure \ref{fig:player_variance}).  Although measurement error is usually not a big concern in basketball analytics, scorekeepers and referees can introduce bias \citep{van2017adjusting, price2010racial}.  We also emphasize that basketball is a team sport, and thus metrics for individual performance are impacted by the abilities of their teammates.  Since observed metrics are influenced by many factors, when devising a method targeted at a specific quantity, the first step is to clearly distinguish the relevant sources of variability from the irrelevant nuisance variability.

% Cite Figure

To characterize the effect of these sources of variability on existing basketball metrics, \citet{franks2016meta} proposed a set of three ``meta-metrics": 1) \emph{discrimination}, which quantifies the extent to which a metric actually reflects true differences between player skill rather than chance variation 2) \emph{stability}, which characterizes how a player-metric evolves over time due to development and contextual changes and 3) \emph{independence}, which describes redundancies in the information provided across multiple related metrics.  Arguably, the most useful measures of player performance are metrics that are discriminative and reflect robust measurement of the same (possibly latent) attributes over time.

One of the most important tools for minimizing nuisance variability in characterizing player performance is shrinkage estimation via hierarchical modeling. In their seminal paper, \citet{efron1975data} provide a theoretical justification for hierarchical modeling as an approach for improving estimation in low sample size settings, and demonstrate the utility of shrinkage estimation for estimating batting averages in baseball. Similarly, in basketball, hierarchical modeling is used to leverage commonalities across players by imposing a shared prior on parameters associated with individual performance.  We repeatedly return to these ideas about sources of variability and the importance of hierarchical modeling below. 

% While issues regarding sample size have long been understood by the casual sports fan, there are principles methods for deal ..  

% throughout our discussion of player performance.

 \subsection{General skill}
 \label{sec:general_skill}

One of the most common questions across all sports is ``who is the best player?'' This question takes many forms, ranging from who is the ``most valuable'' in MVP discussions, to who contributes the most to helping his or her team win, to who puts up the most impressive numbers. Some of the most popular metrics for quantifying player-value are constructed using only box score data. These include Hollinger's PER \citep{kubatko2007starting}, Wins Above Replacement Player (WARP)~\citep{pelton}, Berri's quantification of a player's win production~\citep{berri1999most}, Box Plus-Minus (BPM), and Value Over Replacement Player (VORP)~\citep{myers}.  These metrics are particularly useful for evaluating historical player value for players who pre-dated play-by-play and tracking data.  In this review, we focus our discussion on more modern approaches like the regression-based models for play-by-play data and metrics based on tracking data. 

\subsubsection{Regression-based approaches}
\label{reg-section}

One of the first and simplest play-by-play metrics aimed at quantifying player value is known as ``plus-minus''.  A player's plus-minus is computed by adding all of the points scored by the player's team and subtracting all the points scored against the player's team while that player was in the game. However, plus-minus is particularly sensitive to teammate contributions, since a less-skilled player may commonly share the floor with a more-skilled teammate, thus benefiting from the better teammate's effect on the game. Several regression approaches have been proposed to account for this problem.  \citet{rosenbaum} was one of the first to propose a regression-based approach for quantifying overall player value which he terms adjusted plus-minus, or APM~\citep{rosenbaum}. In the APM model, Rosenbaum posits that

\begin{equation}
\label{eqn:pm}
    D_i = \beta_0 + \sum_{p=1}^P\beta_p x_{ip} + \epsilon_i
\end{equation}

\noindent where $D_i$ is 100 times the difference in points between the home and away teams in stint $i$; $x_{ip} \in \lbrace 1, -1, 0 \rbrace $ indicates whether player $p$ is at home, away, or not playing, respectively; and $\epsilon$ is the residual. Each stint is a stretch of time without substitutions. Rosenbaum also develops statistical plus-minus and overall plus-minus which reduce some of the noise in pure adjusted plus-minus~\citep{rosenbaum}. However, the major challenge with APM and related methods is multicollinearity: when groups of players are typically on the court at the same time, we do not have enough data to accurately distinguish their individual contributions using plus-minus data alone.  As a consequence, inferred regression coefficients, $\hat \beta_p$, typically have very large variance and are not reliably informative about player value.  

APM can be improved by adding a penalty via ridge regression~\citep{sill2010improved}. The penalization framework, known as regularized APM, or RAPM, reduces the variance of resulting estimates by biasing the coefficients toward zero~\citep{jacobs_2017}. In RAPM, $\hat \beta$ is the vector which minimizes the following expression 

\begin{equation}
     \mathbf{\hat \beta} = \underset{\beta}{\argmin }(\mathbf{D} - \mathbf{X} \beta)^T (\mathbf{D} - \mathbf{X}\beta) + \lambda \beta^T 
    \beta 
\end{equation}

\noindent where $\mathbf{D}$ and $\mathbf{X}$ are matrices whose rows correspond to possessions and $\beta$ is the vector of skill-coefficients for all players.  $\lambda \beta^T \beta$ represents a penalty on the magnitude of the coefficients, with $\lambda$ controlling the strength of the penalty. The penalty ensures the existence of a unique solution and  reduces the variance of the inferred coefficients.  Under the ridge regression framework, $\hat \beta = (X^T X + \lambda I)^{-1}X^T D$ with $\lambda$ typically chosen via cross-validation.  An alternative formulation uses the lasso penalty, $\lambda \sum_p |\beta_p|$, instead of the ridge penalty~\citep{omidiran2011pm}, which encourages many players to have an adjusted plus-minus of exactly zero.

Regularization penalties can equivalently be viewed from the Bayesian perspective, where ridge regression estimates are equivalent to the posterior mode when assuming mean-zero Gaussian prior distributions on $\beta_p$ and lasso estimates are equivalent to the posterior mode when assuming mean-zero Laplace prior distributions. Although adding shrinkage priors ensures identifiability and reduces the variance of resulting estimates, regularization is not a panacea: the inferred value of players who often share the court is sensitive to the precise choice of regularization (or prior) used.  As such, careful consideration should be placed on choosing appropriate priors, beyond common defaults like the mean-zero Gaussian or Laplace prior.  More sophisticated informative priors could be used; for example, a prior with right skewness to reflect beliefs about the distribution of player value in the NBA, or player- and position-specific priors which incorporate expert knowledge.  Since coaches give more minutes to players that are perceived to provide the most value, a prior on $\beta_p$ which is a function of playing time could provide less biased estimates than standard regularization techniques, which shrink all player coefficients in exactly the same way. APM estimates can also be improved by incorporating data across multiple seasons, and/or by separately inferring player's defensive and offensive contributions, as explored in \citet{fearnhead2011estimating}.

% https://www.basketball-reference.com/about/bpm.html   

%The regularization approaches by Omidiran and Sill attempt to resolve the issue of collinearity -- in this case, players who commonly share the floor together affecting each other's ratings -- but it does not completely solve the problem of measuring a player's value or contribution to winning. 
% \franks{I think we should say a little more here.  THE fundamental challenge with these methods is dealing with collinearity.  Does regularization solve it?} 

\begin{comment}
discuss regression in terms of apm. This starts with Rosenbaum's APM and regressions there, continues with Sill. Sill adds ridge regression; Omidiran adds lasso. 
Fearnhead is similar to Rosenbaum, but "
The main difference compared with Rosenbaum (2004) is that the author estimates only a combined ability for each player. The model presented here further uses a structured approach to combining information from multiple seasons."
\end{comment}

Several variants and alternatives to the RAPM metrics exist. For example,~\citet{page2007using} use a hierarchical Bayesian regression model to identify a position's contribution to winning games, rather than for evaluating individual players.~\citet{deshpande2016estimating} propose a Bayesian model for estimating each player's effect on the team's chance of winning, where the response variable is the home team's win probability rather than the point spread. Models which explicitly incorporate the effect of teammate interactions are also needed. \citet{piette2011evaluating} propose one  approach based on modeling players as nodes in a network, with edges between  players that shared the court together. Edge weights correspond to a measure of performance for the lineup during their shared time on the court, and a measure of network centrality is used as a proxy for player importance.  An additional review with more detail on possession-based player performance can be found in \citet{engelmann2017possession}.

\begin{comment}
Deshpande, Jensen: "We propose instead to regress the change in the home team’s win probability during a shift onto signed indicators corresponding to the five home team players and five away team players in order to estimate each player’s partial effect on his team’s chances of winning."
This paper usefully explains Rosenbaum's approach too:
"To compute Adjusted Plus-Minus, one first breaks the game into several “shifts,” periods of play between substitutions, and measures both the point differential and total number of possessions in each shift. One then regresses the point differential per 100 possessions from the shift onto indicators corresponding to the ten players on the court."
\end{comment}

\subsubsection{Expected Possession Value}
\label{sec:epv}
The purpose of the Expected Possession Value (EPV) framework, as developed by~\citet{cervone2014multiresolution}, is to infer the expected value of the possession at every moment in time.  Ignoring free throws for simplicity, a possession can take on values $Z_i \in \{0, 2, 3\}$. The EPV at time $t$ in possession $i$ is defined as 
\begin{equation}
\label{eqn:epv}
v_{it}=\mathbb{E}\left[Z_i | X_{i0}, ..., X_{it}\right]    
\end{equation}
\noindent where $X_{i0}, ..., X_{it}$ contain all available covariate information about the game or possession for the first $t$ timestamps of possession $i$.  The EPV framework is quite general and can be applied in a range of contexts, from evaluating strategies to constructing retrospectives on the key points or decisions in a possession. In this review, we focus on its use for player evaluation and provide a brief high-level description of the general framework.

~\citet{cervone2014multiresolution} were the first to propose a tractable multiresolution approach for inferring EPV from optical tracking data in basketball. They model the possession at two separate levels of resolution. The \emph{micro} level includes all spatio-temporal data for the ball and players, as well as annotations of events, like a pass or shot, at all points in time throughout the possession. Transitions from one micro state to another are complex due to the high level of granularity in this representation. The \emph{macro} level represents a coarsening of the raw data into a finite collection of states. The macro state at time $t$, $C_t = C(X_t)$, is the coarsened state of the possession at time $t$ and can be classified into one of three state types: $\mathcal{C}_{poss}, \mathcal{C}_{trans},$ and $\mathcal{C}_{end}.$  The information used to define $C_t$ varies by state type. For example,
$\mathcal{C}_{poss}$ is defined by the ordered triple containing the ID of the player with the ball, the location of the ball in a discretized court region, and an indicator for whether the player has a defender within five feet of him or her. $\mathcal{C}_{trans}$ corresponds to ``transition states'' which are typically very brief in duration, as they include moments when the ball is in the air during a shot, pass, turnover, or immediately prior to a rebound: $\mathcal{C}_{trans} = $\{shot attempt from $c \in \mathcal{C}_{poss}$, pass from $c \in \mathcal{C}_{poss}$ to $c' \in \mathcal{C}_{poss}$, turnover in progress, rebound in progress\}. Finally, $\mathcal{C}_{end}$ corresponds to the end of the possession, and simply encodes how the possession ended and the associated value: a made field goal, worth two or three points, or a missed field goal or a turnover, worth zero points.  Working with  macrotransitions facilitates inference, since the macro states are assumed to be semi-Markov, which means the sequence of new states forms a homogeneous Markov chain~\citep{bornn2017studying}.  

Let $C_t$ be the current state and $\delta_t > t$ be the time that the next non-transition state begins, so that $C_{\delta_t} \notin \mathcal{C}_{trans}$ is the next possession state or end state to occur after $C_t$. If we assume that coarse states after time $\delta_t$ do not depend on the data prior to $\delta_t$, that is 

\begin{equation}
\textrm{for } s>\delta_{t}, P\left(C_s \mid C_{\delta_{t}}, X_{0}, \ldots, X_{t}\right)=P\left(C_{s} | C_{\delta_{t}}\right),
\end{equation}

\noindent then EPV can be defined in terms of macro and micro factors as

\begin{equation}
v_{it}=\sum_{c} \mathbb{E}\left[Z_i | C_{\delta_{t}}=c\right] P\left(C_{\delta_{t}}=c | X_{i0}, \ldots, X_{it}\right)
\end{equation}
\noindent since the coarsened Markov chain is time-homogeneous. $\mathbb{E}\left[Z | C_{\delta_{t}}=c\right]$ is macro only, as it does not depend on the full resolution spatio-temporal data. It can be inferred by estimating the transition probabilities between coarsened-states and then applying standard Markov chain results to compute absorbing probabilities. Inferring macro transition probabilities could be as simple as counting the observed fraction of transitions between states, although model-based approaches would likely improve inference.

The micro models for inferring the next non-transition state (e.g. shot outcome, new possession state, or turnover) given the full resolution data, $P(C_{\delta_{t}}=c | X_{i0}, \ldots, X_{it}),$ are more complex and vary depending on the state-type under consideration.~\citet{cervone2014multiresolution} use log-linear hazard models~\citep[see][]{prentice1979hazard} for modeling both the time of the next major event and the type of event (shot, pass to a new player, or turnover), given the locations of all players and the ball. \citet{sicilia2019deephoops} use a deep learning representation to model these transitions.  The details of each transition model depend on the state type: models for the case in which $C_{\delta_t}$ is a shot attempt or shot outcome are discussed in Sections \ref{sec:shot_efficiency} and \ref{sec:shot_selection}. See~\citet{masheswaran2014three} for a discussion of factors relevant to modeling rebounding and the original EPV papers for a discussion of passing models~\citep{cervone2014multiresolution, bornn2017studying}.
% \Terner{Skip from here...}

% These coarsened states can be represented as $C_t = $

% They consider all information in a possession, including $(x,y)$ coordinates for all ten players and $(x,y,z)$ coordinates for the ball, information about the game situation, and annotations that happen in real time such as a pass or shot attempt. 

% Macrostransitions which correspond to 
% Microtransitions correspond small player movements occuring when major ball movement does not occur and 

%~\citep{cervone2014multiresolution} 

% the time series created by this possession, where $Z$ contains $(x,y)$ coordinates for all 10 players, $(x,y,z)$ coordinates for the ball, information about the game situation, and annotations that happen in real time, such as a pass or shot attempt. 

% % Defining $Z(\omega)$ as a stochastic process creates the natural filtration $\mathcal{F}_{t}^{(Z)}=\sigma\left(\left\{Z_{s}^{-1}: 0 \leq s \leq t\right\}\right),$ 

% % This multiresolution approach allows for the feasible computation of EPV by conditioning on only the current relevant information in the possession -- such as the player with the ball, his location on the court, and the amount of defensive pressure -- rather than the history of the entire possession. Computing EPV in this way not only preserves computational feasibility, but also provides a more informed view of a player's contribution since tracking data contain more information than the shifts used in variants of plus-minus. 

% \Terner{...to here? Can smooth the transition.}

~\citet{cervone2014multiresolution} suggested two metrics for characterizing player ability that can be derived from EPV: Shot Satisfaction (described in Section \ref{sec:shot_selection}) and EPV Added (EPVA), a metric quantifying the overall contribution of a player. EPVA quantifies the value relative to the league average of an offensive player receiving the ball in a similar situation. A player $p$ who possesses the ball starting at time $s$ and ending at time $e$ contributes value $v_{t_e} - v_{t_s}^{r(p)}$ over the league average replacement player, $r(p)$. Thus, the EPVA for player $p$, or EPVA$(p)$, is calculated as the average value that this player brings over the course of all times that player possesses the ball:

\begin{equation}
\text{EPVA(p)} =  \frac{1}{N_p}\sum_{\{t_s, t_e\} \in \mathcal{T}^{p}} v_{t_e} - v_{t_s}^{r(p)}
\end{equation}

\noindent where $N_p$ is the number of games played by $p$, and $\mathcal{T}^{p}$ is the set of starting and ending ball-possession times for $p$ across all games.  Averaging over games, instead of by touches, rewards high-usage players.  Other ways of normalizing EPVA, e.g. by dividing by $|\mathcal{T}^p|$, are also worth exploring.

Unlike RAPM-based methods, which only consider changes in the score and the identities of the players on the court, EPVA leverages the high resolution optical data to characterize the precise value of specific decisions made by the ball carrier throughout the possession.  Although this approach is powerful, it still has some crucial limitations for evaluating overall player value. The first is that EPVA measures the value added by a player only when that player touches the ball. As such, specialists, like three point shooting experts, tend to have high EPVA because they most often receive the ball in situations in which they are uniquely suited to add value.  However, many players around the NBA add significant value by setting screens or making cuts which draw defenders away from the ball. These actions are hard to measure and thus not included in the original EPVA metric proposed by \citet{cervone2014multiresolution}.  In future work, some of these effects could be captured by identifying appropriate ways to measure a player's ``gravity''~\citep{visualizegravity} or through new tools which  classify important off-ball actions.  Finally, EPVA only represents contributions on the offensive side of the ball and ignores a player's defensive prowess; as noted in Section~\ref{defensive ability}, a defensive version of EPVA would also be valuable. 

In contrast to EPVA, the effects of off-ball actions and defensive ability are implicitly incorporated into RAPM-based metrics.  As such, RAPM remains one of the key metrics for quantifying overall player value.  EPVA, on the other hand, may provide better contextual understanding of how players add value, but a less comprehensive summary of each player's total contribution. A more rigorous comparison between RAPM, EPVA and other metrics for overall ability would be worthwhile.

% Each of these contributions are broadly contained in plus-minus statistics, though they are by no means identifiable. These abilities warrant future work in defining and measuring them more closely.

% It is common to view the locations of events from a player as the result of a heterogeneous spatial point process on the court \citep{cervone2014multiresolution, reich2006spatial, miller2013icml}.  The probability of an event occuring in an area A is ...

% $N_{A} \sim \rm{Pois}\left(\int_{A} \lambda(d A)\right)$

 \subsection{Production curves}
  \label{sec:production_curves}
  
A major component of quantifying player ability involves understanding how ability evolves over a player's career. To predict and describe player ability over time, several methods have been proposed for inferring the so-called ``production curve'' for a player\footnote{Production curves are also referred to as ``player aging curves'' in the literature, although we prefer ``production curves'' because it does not imply that changes in these metrics over time are driven exclusively by age-related factors.}. The goal of a production curve analysis is to provide predictions about the future trajectory of a current player's ability, as well as to characterize similarities in production trajectories across players.  These two goals are intimately related, as the ability to forecast production is driven by assumptions about historical production from players with similar styles and abilities.  
 
Commonly, in a production curve analysis, a continuous measurement of aggregate skill (i.e. RAPM or VORP), denoted $\mathbf Y$ is considered for a particular player at time t:

% , is considered for player, $p$, at various times $t_{pj}$. 

$$Y_{pt} = f_p(t) + \epsilon_{pt}$$
\noindent where $f_p$ describes player $p$'s ability as a function of time, $t$, and $\epsilon_{pt}$ reflects irreducible errors which are uncorrelated over time, e.g. due to unobserved factors like minor injury, illness and chance variation. Athletes not only exhibit different career trajectories, but their careers occur at different ages, can be interrupted by injuries, and include different amounts of playing time.  As such, the statistical challenge in production curve analysis is to infer smooth trajectories $f_p(t)$ from sparse irregular observations of $Y_{pt}$ across players \citep{wakim2014functional}.  

There are two common approaches to modeling production curves: 1) Bayesian hierarchical modeling and 2) methods based on functional data analysis and clustering.  In the Bayesian hierarchical paradigm, ~\citet{berry1999bridging} developed a flexible hierarchical aging model to compare player abilities across different eras in three sports: hockey, golf, and baseball.  Although not explored in their paper, their framework can be applied to basketball to account for player-specific development and age-related declines in performance. ~\citet{page2013effect} apply a similar hierarchical method based on Gaussian Process regressions to infer how production evolves across different basketball positions.  They find that production varies across player type and show that point guards (i.e. agile ball-handlers) generally spend a longer fraction of their career improving than other player types.  \citet{vaci2019large} also use a Bayesian hierarchical modeling with distinct parametric curves to describe trajectories before and after peak-performance.  They assume pre-peak performance reflects development whereas post-peak performance is driven by aging. Their findings suggest that athletes which develop more quickly also exhibit slower age-related declines, an observation which  does not appear to depend on position.  

In contrast to hierarchical Bayesian models, \citet{wakim2014functional} discuss how the tools of functional data analysis can be used to model production curves.  In particular, functional principal components metrics can be used in an unsupervised fashion to identify clusters of players with similar trajectories. Others have explicitly incorporated notions of player similarity into functional models of production.  In this framework, the production curve for any player $p$ is then expressed as a linear combination of the production curves from a set of similar players: $f_p(t) \approx \sum_{k \neq p} \alpha_{pk} f_k(t)$.  For example, in their RAPTOR player rating system, \url{fivethirtyeight.com} uses a nearest neighbor algorithm to characterize similarity between players~\citep{natesilver538_2015, natesilver538_2019}. The production curve for each player is an average of historical production curves from a distinct set of the most similar athletes.   A related approach, proposed by \citet{vinue2019forecasting}, employs the method of archetypoids \citep{vinue2015archetypoids}.  Loosely speaking, the archetypoids consist of a small set of players, $\mathcal{A}$, that represent the vertices in the convex hull of production curves.  Different from the RAPTOR approach, each player's production curve is represented as a convex combination of curves from the \emph{same set} of archetypes, that is, $\alpha_{pk} = 0 \; \forall \ k \notin \mathcal{A}$.  

One often unaddressed challenge is that athlete playing time varies across games and seasons, which means sampling variability is non-constant.  Whenever possible, this heteroskedasticity in the observed outcomes should be incorporated into the inference, either by appropriately controlling for minutes played or by using other relevant notions of exposure, like possessions or attempts.  

Finally, although the precise goals of these production curve analyses differ, most current analyses focus on aggregate skill. More work is needed to capture what latent player attributes drive these observed changes in aggregate production over time.  Models which jointly infer how distinct measures of athleticism and skill co-evolve, or models which account for changes in team quality and adjust for injury, could lead to further insight about player ability, development, and aging (see Figure \ref{fig:player_variance}).  In the next sections we mostly ignore how performance evolves over time, but focus on quantifying some specific aspects of basketball ability, including shot making and defense.

 \subsection{Shot modeling}
\label{sec:shooting}

Arguably the most salient aspect of player performance is the ability to score.  There are two key factors which drive scoring ability: the ability to selectively identify the highest value scoring options (shot selection) and the ability to make a shot, conditioned on an attempt (shot efficiency). A player's shot attempts and his or her ability to make them are typically related.  In \emph{Basketball on Paper}, Dean Oliver proposes the notion of a ``skill curve,'' which roughly reflects the inverse relationship between a player's shot volume and shot efficiency \citep{oliver2004basketball, skinner2010price, goldman2011allocative}. Goldsberry and others gain further insight into shooting behavior by visualizing how both player shot selection and efficiency vary spatially with a so-called ``shot chart.'' (See \citet{goldsberry2012courtvision} and \citet{goldsberry2019sprawlball} for examples.) Below, we discuss statistical models for inferring how both shot selection and shot efficiency vary across players, over space, and in defensive contexts.

% Studying the factors that affect shooting behavior can provide deeper insight into performance.  

\subsubsection{Shot efficiency} 
\label{sec:shot_efficiency}
%%% Say something about eFG%?

Raw FG\% is usually a poor measure for the shooting ability of an athlete because chance variability can obscure true differences between players.  This is especially true when conditioning on additional contextual information like shot location or shot type, where sample sizes are especially small.  For example, \citet{franks2016meta} show that the majority of observed differences in 3PT\% are due to sampling variability rather than true differences in ability, and thus is a poor metric for player discrimination.  They demonstrate how these issues can be mitigated by using hierarchical models which shrink empirical estimates toward more reasonable prior means.  These shrunken estimates are both more discriminative and more stable than the raw percentages.  

\begin{figure}
% \begin{wrapfigure}{r}{0.75\textwidth}
    \centering
        \begin{subfigure}[b]{0.4\textwidth}
    \includegraphics[width=\textwidth]{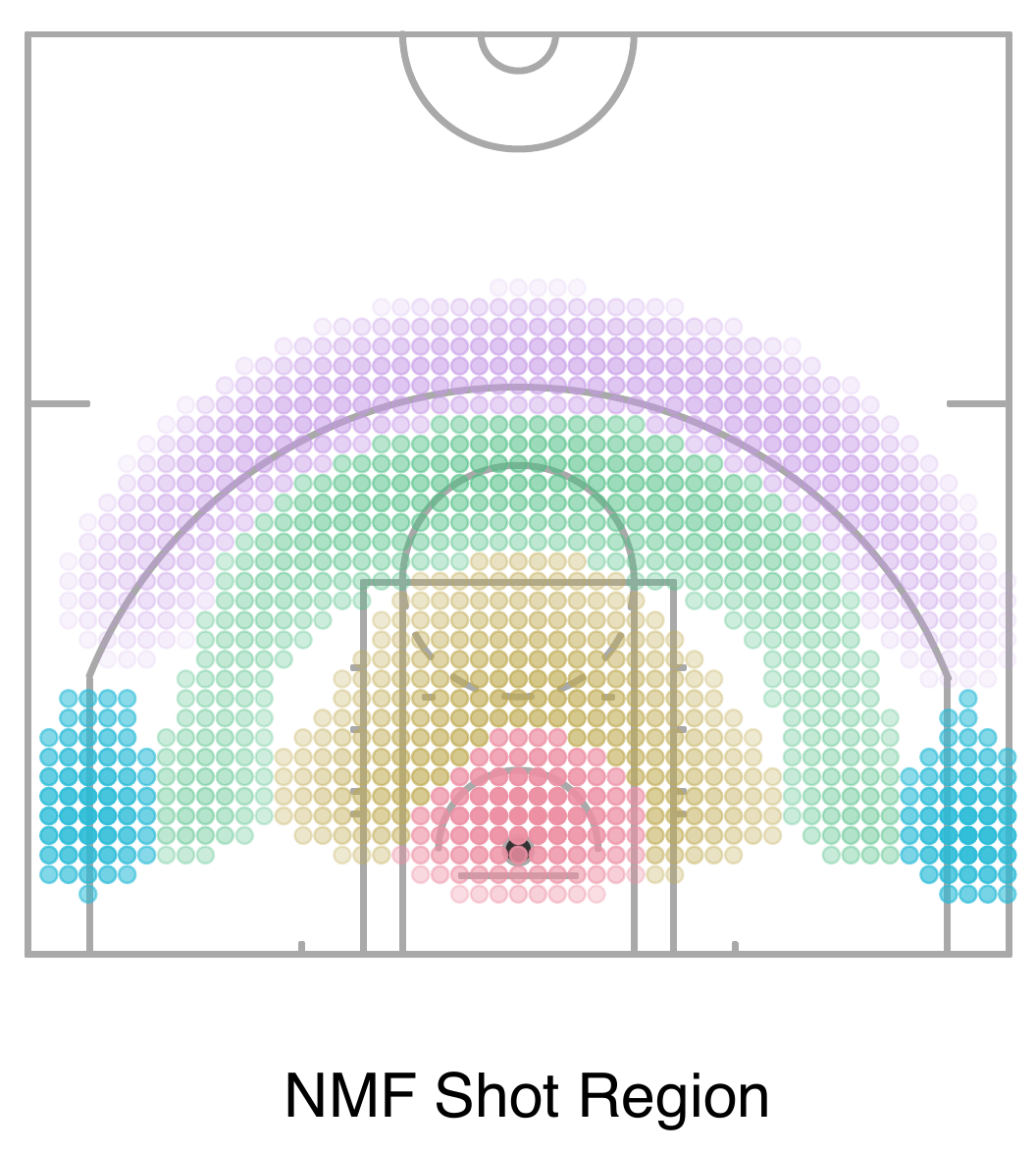}
    \end{subfigure}
    ~~
    \begin{subfigure}[b]{0.55 \textwidth}
    \includegraphics[width=\textwidth]{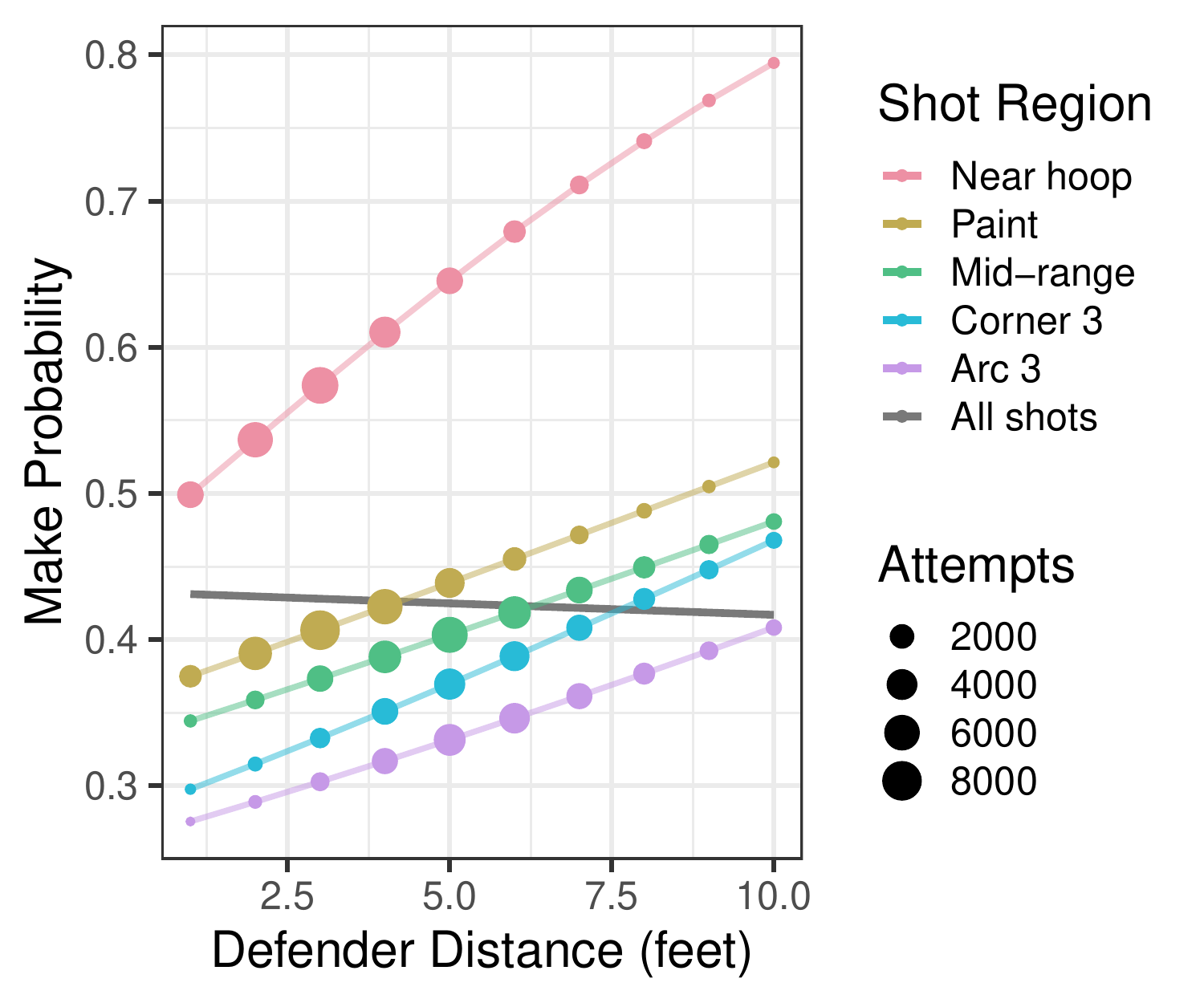}
    \end{subfigure}

    \caption{Left) The five highest-volume shot regions, inferred using the NMF method proposed by \citet{miller2013icml}. Right) Fitted values in a logistic regression of shot outcome given defender distance and NMF shot region from over 115,000 shot attempts in the 2014-2015 NBA season \citep{franks2015characterizing, simpsons_personal}.   The make probability increases approximately linearly with increasing defender distance in all shot locations.  The number of observed shots at each binned defender distance is indicated by the point size.  Remarkably, when ignoring shot region, the coefficient of defender distance has a slightly \emph{negative} coefficient, indicating that the probability of making a shot increases slightly with the closeness of the defender (gray line). This effect, which occurs because defender distance is also dependent on shot region, is an example of a ``reversal paradox''~\citep{tu2008simpson} and highlights the importance of accounting for spatial context in basketball. It also demonstrates the danger of making causal interpretations without carefully considering the role of confounding variables. }
    \label{fig:simpsons}
 \end{figure}

With the emergence of tracking data, hierarchical models have been developed which target increasingly context-specific estimands. \citet{franks2015characterizing} and \citet{cervone2014multiresolution} propose similar hierarchical logistic regression models for estimating the probability of making a shot given the shooter identity, defender distance, and shot location.  In their models, they posit the logistic regression model 
\begin{equation}
E[Y_{ip} \mid \ell_{ip}, X_{ijp}] = \textrm{logit}^{-1} \big( \alpha_{\ell_i,p} + \sum_{j=1}^J \beta_{j} X_{ij} \big)
\end{equation} where $Y_{ip}$ is the outcome of the $i$th shot by player $p$ given $J$ covariates $X_{ij}$ (i.e. defender distance) and $\alpha_{{\ell_i}, p}$ is a spatial random effect describing the baseline shot-making ability of player $p$ in location $\ell_i$. As shown in Figure \ref{fig:simpsons}, accounting for spatial context is crucial for understanding defensive impact on shot making.
Given high resolution data, more complex hierarchical models which capture similarities across players and space are needed to reduce the variance of resulting estimators.  Franks et al. propose a conditional autoregressive (CAR) prior distribution for $\alpha_{\ell_i,p}$ to describe similarity in shot efficiencies between players.  The CAR prior is simply a multivariate normal prior distribution over player coefficients with a structured covariance matrix.  The prior covariance matrix is structured to shrink the coefficients of players with low attempts in a given  region toward the FG\%s of players with similar styles and skills.  The covariance is constructed from a nearest-neighbor similarity network on players with similar shooting preferences.  These prior distributions improve out-of-sample predictions for shot outcomes, especially for players with fewer attempts. To model the spatial random effects, they represent a smoothed spatial field as a linear combination of functional bases following a matrix factorization approach proposed by \citet{miller2013icml} and discussed in more detail in Section \ref{sec:shot_selection}.  

More recently, models which incorporate the full 3-dimensional trajectories of the ball have been proposed to further improve estimates of shot ability. Data from SportVU, Second Spectrum, NOAH, or RSPCT include the location of the ball in space as it approaches the hoop, including left/right accuracy and the depth of the ball once it enters the hoop.~\citet{marty2017data} and~\citet{marty2018high} use ball tracking data from over 20 million attempts taken by athletes ranging from high school to the NBA.  From their analyses, \citet{marty2018high} and \citet{daly2019rao} show that the optimal entry location is about 2 inches beyond the center of the basket, at an entry angle of about $45^{\circ}$.

Importantly, this trajectory information can be used to improve estimates of shooter ability from a limited number of shots. \citet{daly2019rao} use trajectory data and a technique known as Rao-Blackwellization to generate lower error estimates of shooting skill. In this context, the Rao-Blackwell theorem implies that one can achieve lower variance estimates of the sample frequency of made shots by conditioning on sufficient statistics; here, the probability of making the shot.  Instead of taking the field goal percentage as $\hat \theta_{FG} = \sum Y_{i} / n$, they infer the percentage as $\hat \theta_{FG\text{-}RB} = \sum p_{i} / n$, where $p_i = E[Y_i \mid X]$ is the inferred probability that shot $i$ goes in, as inferred from trajectory data $X$. The shot outcome is not a deterministic function of the observed trajectory information due to the limited precision of spatial data and the effect of unmeasured factors, like ball spin.  They estimate the make probabilities, $p_i$, from the ball entry location and angle using a logistic regression.

~\citet{daly2019rao} demonstrate that Rao-Blackwellized estimates are better at predicting end-of-season three point percentages from limited data than empirical make percentages.  They also integrate the RB approach into a hierarchical model to achieve further variance reduction. In a follow-up paper, they focus on the effect that defenders have on shot trajectories~\citep{bornn2019using}. Unsurprisingly, they demonstrate an increase in the variance of shot depth, left-right location, and entry angle for highly contested shots, but they also show that players are typically biased toward short-arming when heavily defended.

% shooting percentage caused by contesting shots may be attributed to shooters biasing their shots shorter when confronted with tight defense

% \end{wrapfigure}

\subsubsection{Shot selection}
\label{sec:shot_selection}

Where and how a player decides to shoot is also important for determining one's scoring ability.  Player shot selection is driven by a variety of factors including individual ability, teammate ability, and strategy~\citep{goldman2013live}.  For example, \citet{alferink2009generality} study the psychology of shot selection and how the positive ``reward'' of shot making affects the frequency of attempted shot types. The log relative frequency of two-point shot attempts to three-point shot attempts is approximiately linear in the log relative frequency of the player's ability to make those shots, a relationship known to psychologists as the generalized matching law~\citep{poling2011matching}. \citet{neiman2011reinforcement} study this phenomenon from a reinforcement learning perspective and demonstrate that a previous made three point shot increases the probability of a future three point attempt. Shot selection is also driven by situational factors, strategy, and the ability of a player's teammates. \citet{zuccolotto2018big} use nonparametric regression to infer how shot selection varies as a function of the shot clock and score differential, whereas \citet{goldsberry2019sprawlball} discusses the broader strategic shift toward high volume three point shooting in the NBA. 

The availability of high-resolution spatial data has spurred the creation of new methods to describe shot selection.~\citet{miller2013icml} use a non-negative matrix factorization (NMF) of player-specific shot patterns across all players in the NBA to derive a low dimensional representation of a pre-specified number of approximiately disjoint shot regions.  These identified regions correspond to interpretable shot locations, including three-point shot types and mid-range shots, and can even reflect left/right bias due to handedness.  See Figure \ref{fig:simpsons} for the results of a five-factor NMF decomposition.  With the inferred representation, each player's shooting preferences can be approximated as a linear combination of the canonical shot ``bases.''  The player-specific coefficients from the NMF decomposition can be used as a lower dimensional characterization of the shooting style of that player \citep{bornn2017studying}.  

While the NMF approach can generate useful summaries of player shooting styles, it incorporates neither contextual information, like defender distance, nor hierarchical structure to reduce the variance of inferred shot selection estimates.  As such, hierarchical spatial models for shot data, which allow for spatially varying effects of covariates, are warranted  \citep{reich2006spatial, franks2015characterizing}.  \citet{franks2015characterizing} use a hierarchical multinomial logistic regression to predict who will attempt a shot and where the attempt will occur given defensive matchup information.  They consider a 26-outcome multinomial model, where the outcomes correspond to shot attempts by one of the five offensive players in any of five shot regions, with regions determined \textit{a priori} using the NMF factorization.  The last outcome corresponds to a possession that does not lead to a shot attempt. Let $\mathcal{S}(p, b)$ be an indicator for a shot by player $p$ in region $b$. The shot attempt probabilities are modeled as

%  $$p\left(\mathcal{S}_{n}(k, b)=1 | \alpha, Z_{n}\right)=\frac{\exp \left(\alpha_{k b}+\sum_{j=1}^{5} Z_{n}(j, k) \beta_{j b}\right)}{1+\sum_{m b} \exp \left(\alpha_{k b}+\sum_{j=1}^{5} Z_{n}(j, k) \beta_{j b}\right)}$$
\begin{equation}
\label{eqn:shot_sel}
E[\mathcal{S}(p, b) \mid \ell_{ip}, X_{ip}] = \frac{\exp \left(\alpha_{p b}+\sum_{j=1}^{5} F_{n}(j, p) \beta_{j b}\right)}{1+\sum_{\tilde p,\tilde b} \exp \left(\alpha_{\tilde p \tilde b}+\sum_{j=1}^{5} F_{n}(j, \tilde p) \beta_{j \tilde b}\right)}
% \textrm{Softmax}(\alpha_{p b}+\sum_{j=1}^{5} F(j, p) \beta_{j b}) 
\end{equation}

\noindent where $\alpha_{pb}$ is the propensity of the player to shoot from region $b$, and $F(j, p)$ is the fraction of time in the possession that player $p$ was guarded by defender $j$.  Shrinkage priors are again used for the coefficients based on player similarity.  $\beta_{jb}$ accounts for the effect of defender $j$ on offensive player $p$'s shooting habits (see Section \ref{defensive ability}).

% \citep{ervculj2015basketball} differences in shot type across leagues

% Like \citep{bornn2019using}, \citep{zuccolotto2018big} assess how shooting is affected high pressure situations.

Beyond simply describing the shooting style of a player, we can also assess the degree to which players attempt high value shots.  \citet{chang2014quantifying} define effective shot quality (ESQ) in terms of the league-average expected value of a shot given the shot location and defender distance.~\citet{shortridge2014creating} similarly characterize how expected points per shot (EPPS) varies spatially.  These metrics are useful for determining whether a player is taking shots that are high or low value relative to some baseline, i.e., the league average player. 

\citet{cervonepointwise} and \citet{cervone2014multiresolution} use the EPV framework (Section \ref{sec:epv}) to develop a more sophisticated measure of shot quality termed ``shot satisfaction''. Shot satisfaction incorporates both offensive and defensive contexts, including shooter identity and all player locations and abilities, at the moment of the shot.  The ``satisfaction'' of a shot is defined as the conditional expectation of the possession value at the moment the shot is taken, $\nu_{it}$, minus the expected value of the possession conditional on a counterfactual in which the player did not shoot, but passed or dribbled instead.  The shot satisfaction for player $p$ is then defined as the average satisfaction, averaging over all shots attempted by the player:

$$\textrm{Satis}(p)=\frac{1}{\left|\mathcal{T}_{\textrm{shot }}^{p}\right|} \sum_{(i, t) \in \mathcal{T}_{\textrm{shot }}^{p}} \left(v_{it}-\mathbb{E}\left[Z_i | X_{it}, C_{t} \textrm{ is a non-shooting state} \right]\right)$$

\noindent where $\mathcal{T}_{\textrm{shot }}^{p}$ is the set of all possessions and times at which a player $p$ took a shot, $Z_i$ is the point value of possession $i$, $X_{it}$ corresponds to the state of the game at time $t$ (player locations, shot clock, etc) and $C_t$ is a non-shooting macro-state. $\nu_t$ is the inferred EPV of the possession at time $t$ as defined in Equation \ref{eqn:epv}.  Satisfaction is low if the shooter has poor shooting ability, takes difficult shots, or if the shooter has teammates who are better scorers.  As such, unlike other metrics, shot satisfaction measures an individual's decision making and implicitly accounts for the shooting ability of both the shooter \emph{and} the ability of their teammates.  However, since shot satisfaction only averages differential value over the set $\mathcal{T}_{\textrm{shot}}^{p}$, it does not account for situations in which the player passes up a high-value shot.  Additionally, although shot satisfaction is aggregated over all shots, exploring spatial variability in shot satisfaction would be an interesting extension.

\subsubsection{The hot hand} 
\label{hothand}

One of the most well-known and debated questions in basketball analytics is about the existence of the so-called ``hot-hand''.  At a high level, a player is said to have a ``hot hand'' if the conditional probability of making a shot increases given a prior sequence of makes.  Alternatively, given $k$ previous shot makes, the hot hand effect is negligible if $E[Y_{p,t}|Y_{p, t-1}=1, ..., Y_{p, t-k}=1, X_t] \approx E[Y_{p,t}| X_t]$ where $Y_{p, t}$ is the outcome of the $t$th shot by player $p$ and $X_t$ represents contextual information at time $t$ (e.g. shot type or defender distance).  In their seminal paper,~\citet{gilovich1985hot} argued that the hot hand effect is negligible. Instead, they claim streaks of made shots arising by chance are misinterpreted by fans and players \textit{ex post facto} as arising from a short-term improvement in ability. Extensive research following the original paper has found modest, but sometimes conflicting, evidence for the hot hand~\citep[e.g.][]{bar2006twenty, yaari2011hot,hothand93online}.

% Extensive research following the original paper has found modest, but sometimes conflicting, evidence for the hot hand~\citep{bar2006twenty, yaari2011hot}.  For example, \citet{arkes2010revisiting} use hierarchical modeling to estimate player specific conditional expectations based on free throw data and find that on average, ``hot'' players have a 2.9\% higher chance of making their shots. \citet{csapo2014hand} find no evidence for the hot hand, but argue that defenders react to a long streak of makes by increasing the defensive intensity on the streaking player. 

Amazingly, 30 years after the original paper,~\citet{miller2015surprised} demonstrated the existence of a bias in the estimators used in the original and most subsequent hot hand analyses.  The bias, which attenuates estimates of the hot hand effect, arises due to the way in which shot sequences are selected and is closely related to the infamous Monty Hall problem~\citep{sciam, miller2017bridge}.  After correcting for this bias, they estimate that there is an 11\% increase in the probability of making a three point shot given a streak of previous makes, a significantly larger hot-hand effect than had been previously reported. 
% although the practical implications of this effect are some unclear.

% https://www.scientificamerican.com/article/momentum-isnt-magic-vindicating-the-hot-hand-with-the-mathematics-of-streaks/

% The question is not about whether shot making is correlated over time (after all, shot making ability clearly changes over time due to development, fitness and aging), but rather, what the reasonable \emph{timescales} over which we expect meaningful correlation in efficiency (cite gelman blog)?   

% While the original hot hand paper focused on the psychology of the "misperception of chance" events, it is important to consider 

Relatedly,~\citet{stone2012measurement} describes the effects of a form of ``measurement error'' on hot hand estimates, arguing that it is more appropriate to condition on the \emph{probabilities} of previous makes, $E\left[Y_{p,t}|E[Y_{p, t-1}], ... E[Y_{p, t-k}], X_t\right]$, rather than observed makes and misses themselves -- a subtle but important distinction. From this perspective, the work of \citet{marty2018high} and \citet{daly2019rao} on the use of ball tracking data to improve estimates of shot ability could provide fruitful views on the hot hand phenomenon by exploring autocorrelation in shot trajectories rather than makes and misses. To our knowledge this has not yet been studied.  For a more thorough review and discussion of the extensive work on statistical modeling of streak shooting, see \citet{lackritz2017probability}.  

% Finally, it is worth considering the factors that may be driving the hot hand effect.  Many authors have chosen to focus on controlled settings like free throws, but when analyzing in-game data, much of the autocorrelation in shotmaking could be explained away by  covariates like shot selection and defender distance.  Should it still qualify as the hot hand if FG\% increases when temporarily guarded by poor defenders? The answer to these kinds of questions partly depends on whether we are more interested in studying the psychology of the ``misperception of chance" or in the implications of the hot hand for players, fans, and basketball coaches. 

\subsection{Defensive ability}
\label{defensive ability}

Individual defensive ability is extremely difficult to quantify because 1) defense inherently involves team coordination and 2) there are relatively few box scores statistics related to defense.  Recently, this led Jackie MacMullan, a prominent NBA journalist, to proclaim that ``measuring defense effectively remains the last great frontier in analytics''~\citep{espnmac}. Early attempts at quantifying aggregate defensive impact include Defensive Rating (DRtg), Defensive Box Plus/Minus (DBPM) and Defensive Win Shares, each of which can be computed entirely from box score statistics \citep{oliver2004basketball, bbref_ratings}. DRtg is a metric meant to quantify the ``points allowed'' by an individual while on the court (per 100 possessions).  Defensive Win Shares is a measure of the wins added by the player due to defensive play, and is derived from DRtg.  However, all of these measures are particularly sensitive to teammate performance, and thus are not reliable measures of individual defensive ability.  

%% Describe here

% https://www.basketball-reference.com/about t
% Defensive BPM is simply overall BPM minus offensive BPM. The offensive BPM regression was tuned to minimize weighted squared error on both offensive and defensive RAPM simultaneously.

%With the introduction of tracking data, s
Recent analyses have targeted more specific descriptions of defensive ability by leveraging tracking data, but still face some of the same difficulties. Understanding defense requires as much an understanding about what \emph{does not} happen as what does happen. What shots were not attempted and why? Who \emph{did not} shoot and who was guarding them? \citet{goldsberry2013dwight} were some of the first to use spatial data to characterize the absence of shot outcomes in different contexts.  In one notable example from their work, they demonstrated that when Dwight Howard was on the court, the number of opponent shot attempts in the paint dropped by 10\% (``The Dwight Effect''). 

More refined characterizations of defensive ability require some understanding of the defender's goals.  \citet{franks2015characterizing} take a limited view on defenders' intent by focusing on inferring whom each defender is guarding. Using tracking data, they developed an unsupervised algorithm, i.e., without ground truth matchup data, to identify likely defensive matchups at each moment of a possession. They posited that a defender guarding an offensive player $k$ at time $t$ would be normally distributed about the point $\mu_{t k}=\gamma_{o} O_{t k}+\gamma_{b} B_{t}+\gamma_{h} H$, where $O_t$ is the location of the offensive player, $B_t$ is the location of the ball, and $H$ is the location of the hoop.  They use a Hidden Markov model to infer the weights $\mathbf{\gamma}$ and subsequently the evolution of defensive matchups over time.  They find that the average defender location is about 2/3 of the way between the segment connecting the hoop to the offensive player being guarded, while shading about 10\% of the way toward the ball location.~\citet{keshri2019automatic} extend this model by allowing $\mathbf{\gamma}$ to depend on player identities and court locations for a more accurate characterization of defensive play that also accounts for the ``gravity'' of dominant offensive players.

% Their method characterizes the ``gravity'' of offensive players

Defensive matchup data, as derived from these algorithms, is essential for characterizing the effectiveness of individual defensive play. For example, \citet{franks2015characterizing} use matchup data to describe the ability of individual defenders to both suppress shot attempts and disrupt attempted shots at different locations.  To do so, they include defender identities and defender distance in the shot outcome and shot attempt models described in Sections \ref{sec:shot_efficiency} and \ref{sec:shot_selection}.  Inferred coefficients relate to the ability of a defensive player to either reduce the propensity to make a shot given that it is taken, or to reduce the likelihood that a player attempts a shot in the first place.  

% \Terner{continue..}

These coefficients can be summarized in different ways. For example,~\citet{franks2015characterizing} introduce the defensive analogue of the shot chart by visualizing where on the court defenders reduce shot attempts and affect shot efficiency. They found that in the 2013-2014 season, Kawhi Leonard reduced the percentage of opponent three attempts more than any other perimeter defender; Roy Hibbert, a dominant big that year, faced more shots in the paint than any other player, but also did the most to reduce his opponent's shooting efficiency.  In~\citet{franks2015counterpoints}, matchup information is used to derive a notion of ``points against''-- the number of points scored by offensive players when guarded by a specific defender. Such a metric can be useful in identifying the weak links in a team defense, although this is very sensitive to the skill of the offensive players being guarded.

Ultimately, the best matchup defenders are those who encourage the offensive player to make a low value decision. The EPVA metric discussed in Section \ref{sec:general_skill} characterizes the value of offensive decisions by the ball handler, but a similar defender-centric metric could be derived by focusing on changes in EPV when ball handlers are guarded by a specific defender. Such a metric could be a fruitful direction for future research and provide insight into defenders which affect the game in unique ways. Finally, we note that a truly comprehensive understanding of defensive ability must go beyond matchup defense and incorporate aspects of defensive team strategy, including strategies for zone defense.  Without direct information from teams and coaches, this is an immensely challenging task.  Perhaps some of the methods for characterizing team play discussed in Section \ref{sec:team} could be useful in this regard. An approach which incorporates more domain expertise about team defensive strategy could also improve upon existing methods.

%\citep{le2017data}

% \subsection{Misc}
% Rebounding, assists?

%  \subsection{Health and fitness}

\section{Discussion}
% \Terner{One thing we currently omit from discussion that could be added is deep learning methods. Also, it seems that many teams still do not know how to draft, but we haven't discussed that at all in the paper so it may not be fair to mention it here.}

Basketball is a game with complex spatio-temporal dynamics and strategies.  With the availability of new sources of data, increasing computational capability, and methodological innovation, our ability to characterize these dynamics with statistical and machine learning models is improving.  In line with these trends, we believe that basketball analytics will continue to move away from a focus on box-score based metrics and towards models for inferring (latent) aspects of team and player performance from rich spatio-temporal data.  Structured hierarchical models which incorporate more prior knowledge about basketball and leverage correlations across time and space will continue to be an essential part of disentangling player, team, and chance variation. In addition, deep learning approaches for modeling spatio-temporal and image data will continue to develop into major tools for modeling tracking data.

However, we caution that more data and new methods do not automatically imply more insight.  Figure \ref{fig:simpsons} depicts just one example of the ways in which erroneous conclusions may arise when not controlling for confounding factors related to space, time, strategy, and other relevant contextual information. In that example, we are able to control for the relevant spatial confounder, but in many other cases, the relevant confounders may not be observed.  In particular, strategic and game-theoretic considerations are of immense importance, but are typically unknown.  As a related simple example, when estimating field goal percentage as a function of defender distance, defenders may strategically give more space to the poorest shooters. Without this contextual information, this would make it appear as if defender distance is \emph{negatively} correlated with the probability of making the shot. 

% As one noteworthy example, marginally the probability of making a shot is \emph{not} correlated with the closeness of 
% the defender (Figure \ref{fig:simpsons}).  Of course this not because defense doesn't matter, but rather because  the closeness of the defender is also correlated with shot location.   Interestingly, conditioning on shot location may not be enough to fix this problem!  
% What does statistics and machine learning bring to the table? What are the important issues in the future of basketball analytics?
% \begin{itemize}
%     \item Latent variable modeling (understanding of latent structures and abilities
%     \item Recognizing the role of sampling variability
%     \item Causality / counterfactuals / natural experiments
%     \item Uncertainty quantification! (Regularization to get identifiability in adjusted plus minus metrics)
% \end{itemize}

As such, we believe that causal thinking will be an essential component of the future of basketball analytics, precisely because many of the most important questions in basketball are causal in nature.  These questions involve a comparison between an observed outcome and a counterfactual outcome, or require reasoning about the effects of strategic intervention:  ``What would have happened if the Houston Rockets had not adopted their three point shooting strategy?'' or ``How many games would the Bucks have won in 2018 if Giannis Antetokounmpo were replaced with an `average' player?''  Metrics like Wins Above Replacement Player are ostensibly aimed at answering the latter question, but are not given an explicitly causal treatment.  Tools from causal inference should also help us reason more soundly about questions of extrapolation, identifiability, uncertainty, and confounding, which are all ubiquitous in basketball.  Based on our literature review, this need for causal thinking in sports remains largely unmet: there were few works which explicitly focused on causal and/or game theoretic analyses, with the exception of a handful in basketball \citep{skinner2015optimal, sandholtz2018transition} and in sports more broadly \citep{lopez2016persuaded, yamlost, gauriot2018fooled}.  

Finally, although new high-resolution data has enabled increasingly sophisticated methods to address previously unanswerable questions, many of the richest data sources are not openly available. Progress in statistical and machine learning methods for sports is hindered by the lack of publicly available data. We hope that data providers will consider publicly sharing some historical spatio-temporal tracking data in the near future. We also note that there is potential for enriching partnerships between data providers, professional leagues, and the analytics community. Existing contests hosted by professional leagues, such as the National Football League's ``Big Data Bowl''~\citep[open to all,][]{nfl_football_operations}, and the NBA Hackathon~\citep[by application only,][]{nbahack}, have been very popular. Additional hackathons and open data challenges in basketball would certainly be well-received.

\section*{DISCLOSURE}

Alexander Franks is a consultant for a basketball team in the National Basketball Association. This relationship did not affect the content of this review. Zachary Terner is not aware of any affiliations, memberships, funding, or financial holdings that might be perceived as affecting the objectivity of this review. 

\section*{ACKNOWLEDGMENTS}

The authors thank Luke Bornn, Daniel Cervone, Alexander D'Amour, Michael Lopez, Andrew Miller, Nathan Sandholtz, Hal Stern, and an anonymous reviewer for their useful comments, feedback, and discussions.

\newpage
\section*{SUPPLEMENTARY MATERIAL}
\input{supplement_content.tex}

\section*{LITERATURE\ CITED}
\renewcommand{\section}[2]{}%

%\bibliographystyle{ar-style1.bst}
%\bibliography{references}
% \input{bib.tex}

\bibliographystyle{ar-style1.bst}
\bibliography{references}

\end{document}

%% file: supplement_content.tex
%\title{Supplementary material}

\maketitle

\begin{table}[!h]
\begin{adjustwidth}{-2.2cm}{}
\label{softwaretools}
\begin{tabular}{l|l|l|l}
\textbf{Software} & \textbf{Package Name}            & \textbf{Description}                                                  & \textbf{Citation} \\
\hline
\textbf{R}        & \texttt{BAwiR}                            & Scrapes data from international (non-NBA) leagues                     & \cite{bawir}             \\
                  & \texttt{ncaahoopR}                        & Scrapes NCAA data from \url{ESPN.com}                                      & \cite{ncaahoopR}              \\
                  & \texttt{ballr}                            & Scrapes \url{basketball-reference.com}                                      & \cite{ballr}              \\
                  & \texttt{nbastatR}                         & Scrapes \url{basketball-reference.com} and other sites                      & \cite{nbastatR}             \\
 \hline
\textbf{Python}   & \texttt{nba\_py}                          & Python API for \url{stats.nba.com}                                          & \cite{nba-py}              \\
                  & \texttt{nba-api}                          & Python API for \url{stats.nba.com}                                        & \cite{nba-api}              \\
                  & \texttt{py-ball}                          & Improves on \texttt{nba\_py} and also works for WNBA & \cite{py-ball}              \\
                  & \texttt{Sportsreference}                  & Scrapes professional and NCAA Men's Basketball data                   & \cite{sportsref-python}              \\
                  & \texttt{basketball-reference-web-scraper} & Focuses on professional basketball                                    & \cite{python-web-scraper} \\
                  & Kostya Linou's code from Github & Visualizes games from SportVU logs & \cite{linouk23-python} 
\end{tabular}
%\end{adjustwidth}
\newline 
\caption{ \textbf{List of R and Python packages for collecting and analyzing basketball data.} 
In R, the \texttt{BAwiR} package is unique as it scrapes data from international (non-NBA) leagues ~\cite{bawir}. The \texttt{ncaahoopR} package scrapes NCAA data from ESPN.com \cite{ncaahoopR} and appears to be the only R package that focuses on NCAA data. In Python, \texttt{nba\_py, nba-api,} and \texttt{py\_ball} scrape NBA data though \texttt{py\_ball} also collects WNBA data. The \texttt{basketball-reference-web-scraper} and \texttt{Sportsreference} packages each scrape \url{basketball-reference.com}. There are numerous Github repositories with code for visualizing and analyzing NBA data. One notable repository belongs to Kostya Linou, who has code for visualizing games from SportVU logs.}
\end{adjustwidth}
\end{table}
\newpage

%Several pieces of software have been written specifically for the analysis of basketball data. For R, the \texttt{BAwiR} package is unique as it scrapes data from international (non-NBA) leagues ~\cite{bawir}. The \texttt{ncaahoopR} package scrapes NCAA data from ESPN.com \cite{ncaahoopR} and appears to be the only R package that focuses on NCAA data. The \texttt{ballr} package was written to scrape basketball-reference.com, which \texttt{nbastatR} does in addition to a few other basketball websites  \cite{ballr, nbastatR}. Similar software packages exist for use in Python. The first of these, \texttt{nba\_py}, was written as a Python API for stats.nba.com \cite{nba-py}. The \texttt{nba-api} package can be used for similar purposes \cite{nba-api}. One package, \texttt{py\_ball}, was written to improve upon \texttt{nba\_py} and help fill out the documentation. The \texttt{py\_ball} package also works for WNBA applications ~\cite{py-ball}. Two packages exist for scraping data from sports-reference.com. One focuses on professional basketball and is aptly named The Basketball Reference Web Scraper ~\cite{python-web-scraper}. The \texttt{Sportsreference} package in Python provides similar functionality but allows for the scraping of NCAA Men's Basketball data as well \cite{sportsref-python}. Numerous other software projects using basketball data abound on Github, including one that visualizes games from raw SportVU logs ~\cite{linouk23-python}.}

\fontsize{9}{11}\selectfont

\begin{table}[!h]
\label{sitestable}
\begin{adjustwidth}{-2.2cm}{}
\begin{tabular}{l|l|l|l}{}
\textbf{Site}      & \textbf{Author} & \textbf{Data and information provided}                                                                      & \textbf{Citation} \\
\hline
NBA Stuffer        & Serhat Ugur     & Provides data on team and player rest for daily fantasy                                   & \cite{nbastuffer}              \\
Inpredictable      & Mike Beuoy      & NBA, WNBA data; win probability graphs; clutch shooting                            & \cite{inpredictable}              \\
82 Games           & Roland Beech         &   Simple player ratings and sortable clutch stats & \cite{82games}              \\
Cleaning the Glass & Ben Falk        & Advanced NBA statistics, cap/salary info, prediction contest         & \cite{cleaningtheglass}              \\
Sham Sports        & Mark Deeks      & Cap and salary info; database of \textgreater{}3600 players & \cite{shamsports}              \\
Real GM            & Ryan Hoak         & Trade machine                                                                             & \cite{realgm}              \\
Trade NBA          & Zach Rodriguez         & Trade machine                                                                             & \cite{tradenba}              \\
ESPN Trade Machine & ESPN         & Trade machine                                                                             & \cite{espntrademachine}              \\
NBA Math           & Adam Fromal         & Specialty statistics                                                                      & \cite{nbamath}              \\
NBA Miner          & G. Gunday, A. Karasu  & Specialty statistics                                                                      & \cite{nbaminer}  \\
NBA Tattoos          & Ethan Swan    & Player and team tattoo information                                                                      & \cite{swan_2020}  \\\end{tabular}
%\end{adjustwidth}
\newline 
\caption{\textbf{Specialty basketball sites, many of which come from an article on Basketball Insiders~\citep{basketballinsiders}.} This table contains but a handful of the websites and tools on the internet which cater to devout basketball fans. Most of these focus on NBA data, but contain a wide variety of information: NBAStuffer provides information on team and player rest; Inpredictable includes win probability graphs and clutch shooting statistics; 82games.com has unique data on NBA production by and against positions for each team. NBAMiner also provides a number of basketball analytics sites at this URL: \url{https://nbamath.com/stat-resources/}. A few websites, such as Sham Sports, Real GM, and Trade NBA, focus on salary and cap information to let fans see what trades are possible. Cleaning the Glass includes salary information as well as a bevy of advanced NBA statistics. NBA Tattoos may be the most unique of all listed sites since it includes a player database of tattoo information.}
\end{adjustwidth}
\end{table}
\newpage